 \documentclass{ws-p8-50x6-00}
 \usepackage{graphicx,amssymb,amsmath,latexsym}
 \def\lsim{{\lower1.2ex\hbox{$<$}\atop
         {\lower-.7ex\hbox{$\sim$}}}}
 \def\gsim{{\lower1.2ex\hbox{$>$}\atop
         {\lower-.7ex\hbox{$\sim$}}}}

\begin{document}

 \title{SHORT-TIME CRITICAL DYNAMICS}

 \author{L. SCH{\"U}LKE}

 \address{Fachbereich Physik, Universit{\"a}t Siegen, D-57068 Siegen, Germany\\E-mail:
 schuelke@pollux.physik.uni-siegen.de}


 \maketitle

 \abstracts{ An introductory review to short-time critical dynamics is given. 
 From the scaling relation valid already in the early stage of the evolution of
 a system at or near the critical point, one derives power law behaviour for
 various quantities. By a numerical simulation of the system one can measure
 the critical exponents and, by searching for the best power law behaviour, one
 can determine the critical point. Critical slowing down as well as finite size
 corrections are nearly absent, since the correlation length is still small for
 times far before equilibrium is reached. By measuring the (pseudo) critical
 points it is also possible to distinguish (weak) first-order from second-order
 phase transitions. }

 In this talk I like to give an introductory report about the main features of
 short-time critical dynamics. The topic exists since about ten years ago. It will
 be shown that it is possible to measure the critical temperature as well as all
 the static and dynamic critical exponents already in the short-time regime of
 the evolution of statistical systems, far before the equilibrium is reached.
 This can be done by starting either from an initial system of very high
 temperature, suddenly quenched to the critical temperature, or by starting with
 a completely ordered state, leading to the same results. For a comprehensive
 review we
 refer to references\cite{zhe98} and\cite{luo99c}. It will also be shown that
 short-time critical dynamics can in addition be used as a tool to distinguish
 first-order from second order phase transitions by comparing the critical
 temperature obtained from the different starting conditions.

 \vspace{0.5cm} 

 It has for long been known that statistical system for
 particular values of their coupling show critical behaviour. This behaviour is
 mainly characterised by the fact that the correlation length $\xi_x$ becomes
 infinite as well as the corresponding correlation time $\xi_t$. If we define
 the difference between the coupling and the critical one by $\tau$, we have
 \begin{equation} 
 \xi_x\to\tau^{-\nu}, \qquad \xi_t\to\tau^{-\nu z}; \qquad
 \tau=\frac{K-K_c}{K_c}. 
 \label{e1}
 \end{equation} 
 Other quantities approach zero or
 infinity when $\tau\to0$, e.g., the magnetisation of a spin system 
 behaves as 
 \begin{equation}
 M(\tau)\to(\tau)^{\beta}. 
 \end{equation} 
 The exponents $\nu$, $\beta$ and $z$
 are static rsp. dynamic critical exponents. Mainly due to the large correlation
 length and -time there exists a dynamical scaling form. We do not explicitly
 quote the scaling form here, but refer to it in the subsequent discussion.

 In 1989 Janssen, Schaub and Schmittmann\cite{jan89} have shown, that this
 scenario is not only valid in equilibrium, but {\em a scaling relation is
 already valid in the short-time regime} of the evaluation of a critical system.
 The system should be prepared at a very high temperature, with a small
 magnetisation $m_0$ remaining. It is then suddenly quenched to the critical
 temperature and released to the dynamical evolution of model A. The authors
 show, using renormalisation group analysis, that after a macroscopic small time
 $t_{mic}$ a scaling form is valid. For the $k-th$ moment of
 the magnetisation it reads
 \begin{equation}
 M^{(k)}(t,\tau,L,m_0) = b^{-k\beta/\nu} \;
           M^{(k)}(\;t/b^z,\;b^{1/\nu}\tau,\;L/b,\;b^{x_0}m_0).
 \label{e2}
 \end{equation}
 In this equation $b$ is an arbitrary scaling factor, $t$ is the time, $\tau$
 is defined in Eq.(\ref{e1}), L is the linear lattice size, and $m_0 \gsim 0$
 the initial magnetisation. Except of  this additional last argument the scaling
 form looks exactly like that in equilibrium. This argument $m_0$ gives rise to
 a new, independent critical exponent $x_0$, the scaling dimension of the
 initial magnetisation.

 \vspace{0.5cm} 

 The questions arises how to exploit the scaling relation (\ref{e2}) in order to
 get the desired information about the critical exponents. 
 We investigate the system numerically with
 Monte-Carlo simulation at a temperature at 
 or near the critical temperature $T_c$, measuring the magnetisation and its
 second moment, $M(t)$ and $M^{(2)}(t)$, as well as the autocorrelation
 $A(t)=\langle\sum_i S_i(t)\; S_i(0)\rangle$, where a time unit is defined as a
 complete update of the whole lattice. Since the system is in the early 
 stage of the evolution the correlation length is still 
 small and finite size problems are
 nearly absent. Therefore we generally consider $L$ large enough and skip this
 argument. An occasional check is made to prove this assumption. 
 We now choose the scaling factor $b=t^{1/z}$ so
 that the main $t$-dependence on the
 right is cancelled. Expanding 
 the scaling
 form (\ref{e2}) for $k=1$ 
 with respect to the small quantity $t^{x_0/z}\;m_0$, one obtains
 \begin{equation}
 M(t)\sim m_0\;t^\theta\;F(t^{1/\nu z}\tau)
 =m_0\;t^\theta\;\left(1+at^{1/\nu z}\tau+O(\tau^2)\right),\quad
 \label{e3}
 \end{equation}
where $\theta=(x_0-\beta/\nu)/z$ has been introduced. In most cases $\theta>0$,
i.e., for $\tau=0$ 
the small initial magnetisation {\em increases} 
in the short-time region. Figure~1 shows the initial increase for the
3-dimensional Ising model as an example\cite{jas99}.
For small $\tau\not=0$  we have expanded the function 
$F(t^{1/\nu z}\tau)$. There appear
corrections to the simple power law dependent on the sign of $\tau$. Therefore
simulating the system for values of the coupling $K$ 
in the neighbourhood of the critical point one obtains $M_K(t)$ with
non-perfect
power law behaviour, and the critical point $K_c$ can be obtained by
interpolation. Figure~2 shows a plot of the magnetisation for the
2-dimensional 3-state Potts model for three different values of the
coupling\cite{sch96}.
The curve with the best power law behaviour is found for $K_c=1.0055(8)$,
while the exact value is $1.00505$.

\begin{figure}[h]
\vspace*{-3.5cm}\begin{tabular}{cc}
\hspace*{-1cm}\includegraphics[height=8cm]{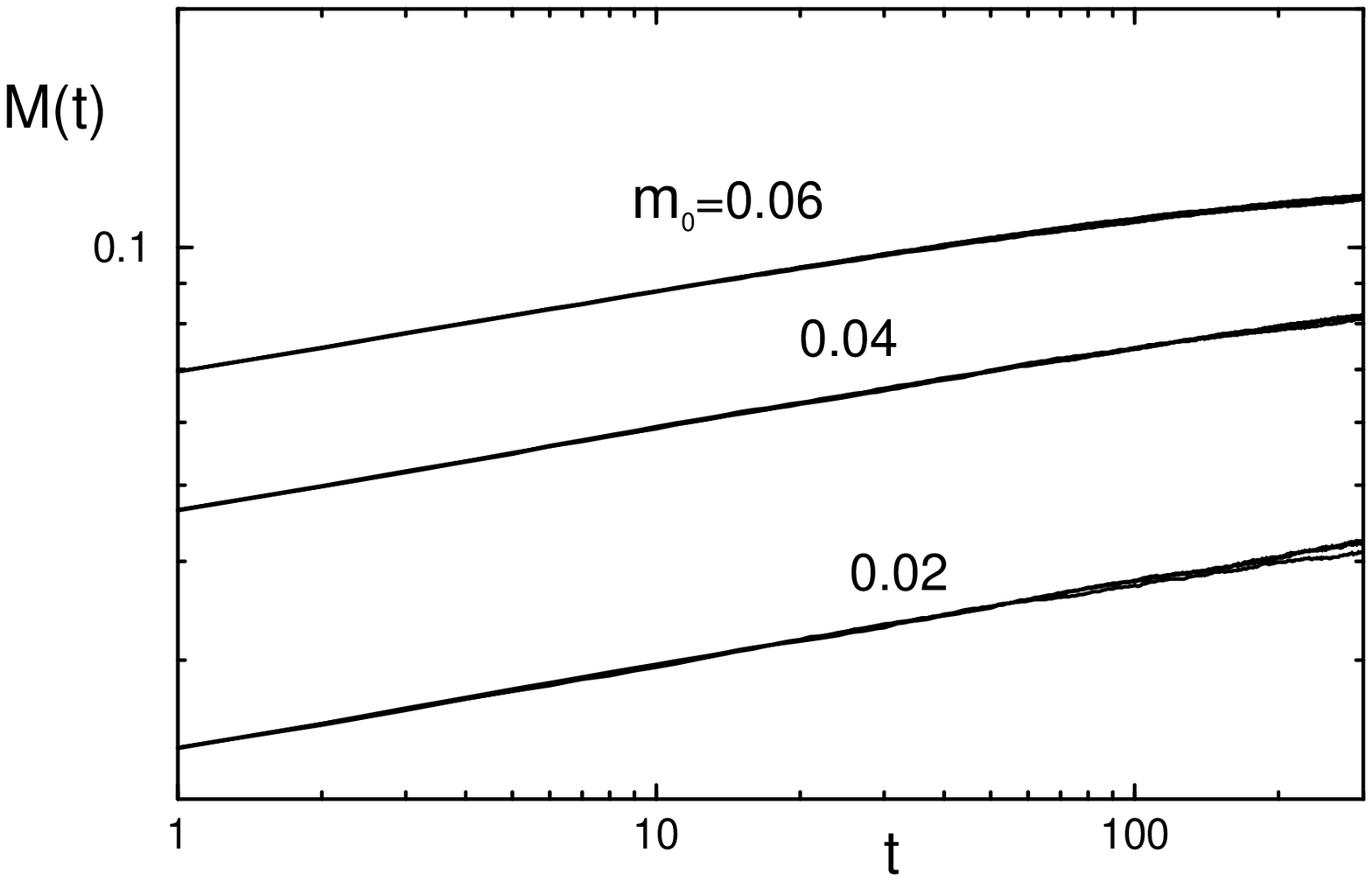} 
&
\raisebox{.5cm}{\includegraphics[height=5cm]{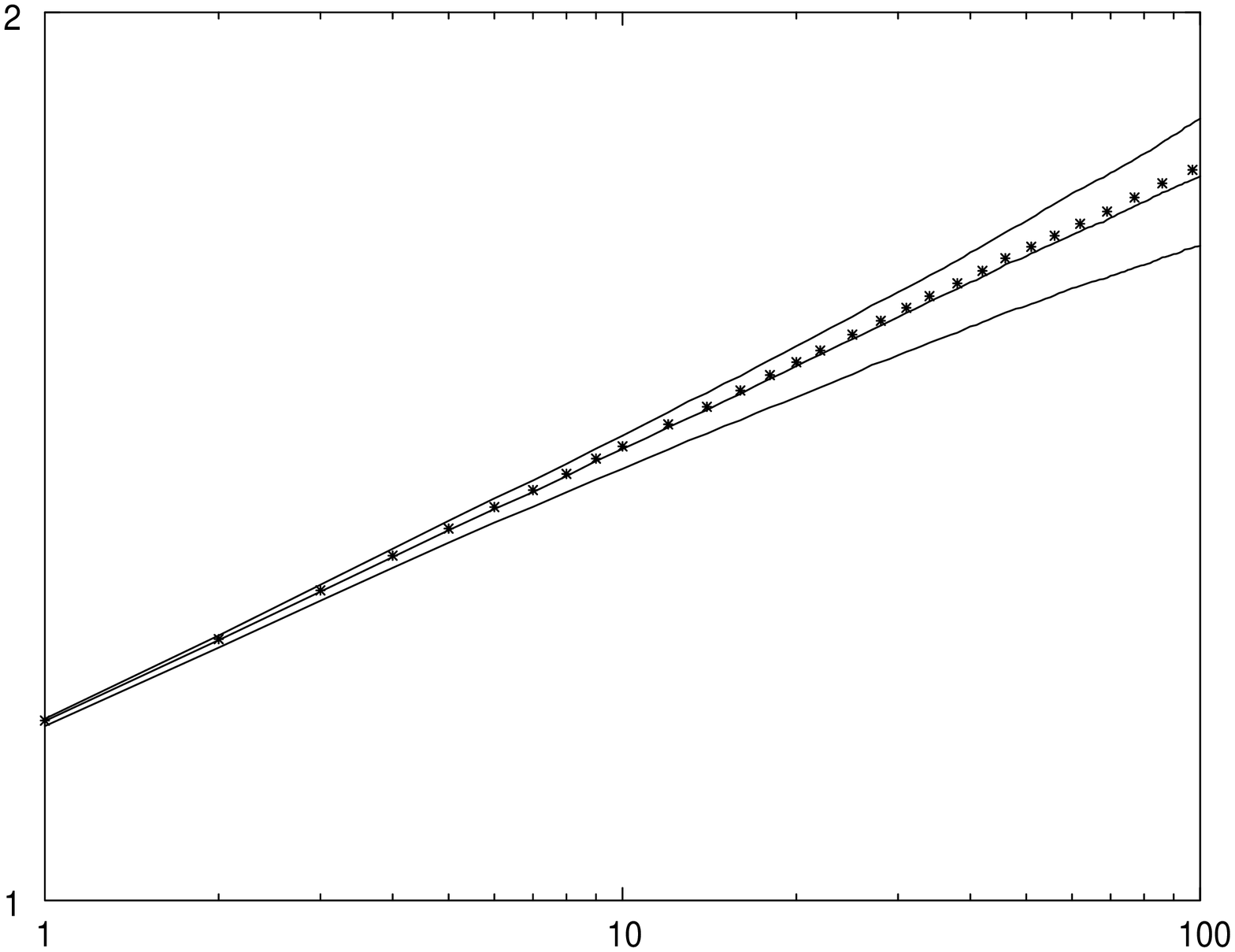}} \\ 
\end{tabular}
\label{f1}
\end{figure}
\vspace*{-1.0cm}
\begin{tabular}{cc}
\hspace*{-1cm}\parbox[l]{5.5cm}{
Figure 1.
Three-dimensional Ising model: Initial increase of the 
magnetisation for
different initial magnetisations $m_0$, plotted vs. time on log-log scale.
$L=128$. Result is $\theta=0.108(2)$}
&
\hspace*{1.cm}\parbox[r]{5.5cm}{Figure 2.
Two-dimensional 3-state Potts model: 
Magnetisation for three different
values of the coupling near the critical point. From above $J=1.00750$,
$1.00505$, and $1.0025$. Dotted curve $J_c=1.0055(8)$}
\\
\end{tabular}

\vspace*{0.4cm}

For the second moment of the magnetisation 
one can deduce \mbox{$M^{(2)}(t)\sim L^{-d}$} 
($d$ = dimension of the system), because the correlation length is still
small in the early time region even at the critical point. Combining this with
the result of the scaling form for $\tau=0$ and $b=t^{1/z}$, 
\mbox{$M^{(2)}(t)\sim t^{-2\beta/\nu z}\;M^{(2)}(1,t^{-1/z}L)$},
one obtains the power law
\begin{equation}
M^{(2)}(t)\sim t^{c_2}\;\qquad c_2=\left(d-2\frac{\beta}{\nu}\right)\frac{1}{z}.
\label{e4}
\end{equation}
An example for the power law behaviour of this quantity is given\cite{jas99}
in Fig.~3.

From the scaling form (\ref{e2}), setting 
again $b=t^{1/z}$, 
one derives for $\partial_\tau\ln M(t,\tau)|_{\tau=0}$ the power
law
\begin{equation}
\partial_\tau\ln M(t,\tau)|_{\tau=0} \sim  t^{\frac{1}{\nu z}}.
\label{e6}
\end{equation}
Approximating numerically the derivative of $M(t)$ involves the difference of
small quantities and is therefore affected more by uncertainties. However,
Fig.~4 shows for the 2-dimensional 3-state Potts model as example
that the data here also, for $t>20$, show perfect power law behaviour. 

Another interesting quantity is the autocorrelation 
\begin{equation}
A(t)=\frac{1}{L^d}\langle\sum_i S_i(t) S_i(0)\rangle.
\label{e5}
\end{equation}
An analysis\cite{jan92} (for $m_0=0$) leads to
\begin{equation}
A(t)\sim t^{c_a}\, \qquad c_a=\frac{d}{z}-\theta.
\label{e5a}
\end{equation}
The plot in Fig.~5 shows a nearly perfect power law. Again the
3-dimensional Ising model has been taken as an example\cite{jas99}.

\begin{figure}[h]
\vspace*{-1cm} 
\begin{tabular}{cc}
\hspace*{-0.5cm}\includegraphics[height=4.5cm]{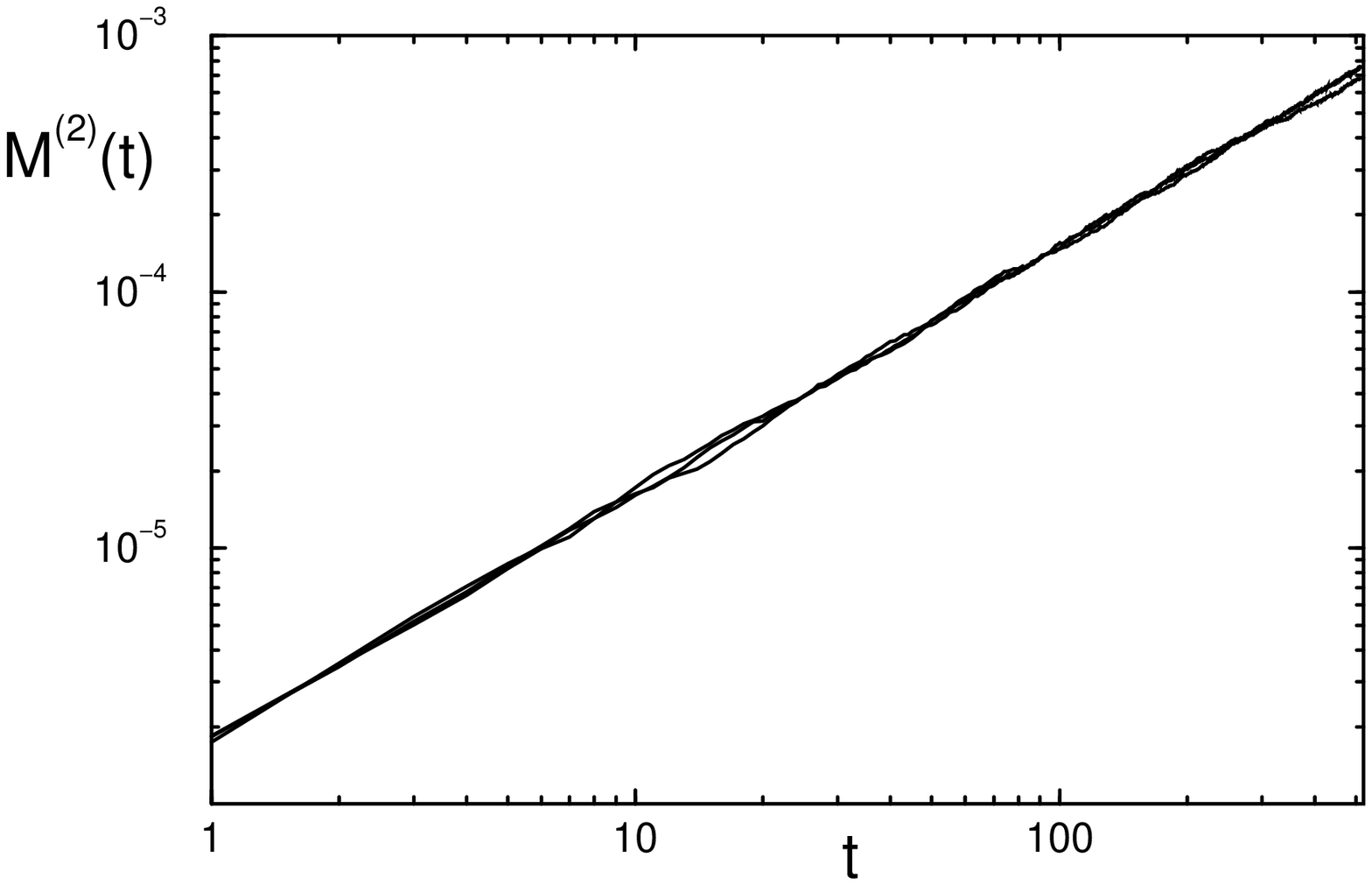} 
&
\hspace*{-.0cm}\raisebox{.5cm}{\includegraphics[height=5.0cm]{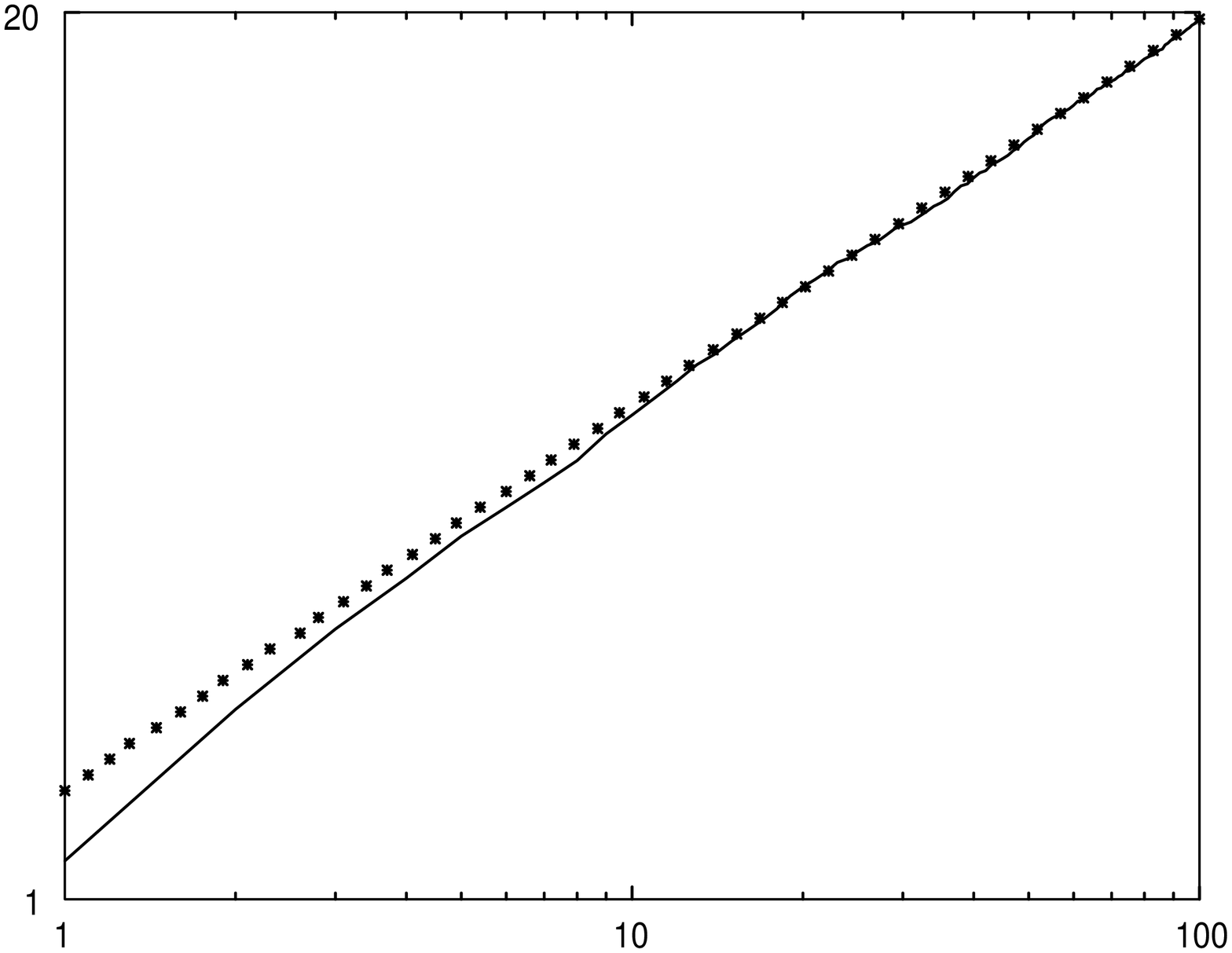}} \\ 
\end{tabular}
\end{figure}

\vspace*{-1cm}
\begin{tabular}{cc}
\hspace*{-1cm}{\parbox[l]{5.7cm}{
Figure 3.
Three-dimensional Ising model:  Second moment of the 
magnetisation, plotted vs. time on log-log scale.
$L=128$. From the slope one gets $c_2=0.970(11)$}}
&
\hspace*{0.5cm}\raisebox{-0.2cm}{\parbox[r]{5.7cm}{Figure 4.
Two-dimensional 3-state Potts model: 
$\partial_\tau\ln M(t,\tau)$,  plotted vs. time on log-log scale. 
Dots represent the straight line fitted to the curve within the time interval
$[20\cdots100]$}}
\\
\end{tabular}

\vspace*{0.2cm}

\noindent Care has to be taken to exclude the time region $t<t_{mic}$, 
as well as the
large time region where deviations of the the power law behaviour occur 
due to finite size effects or due to too large 
fluctuations. An example for a
more detailed analysis of the straight line (on log-log plot) in Fig.~5
is given in Fig.~6.
The individual slope in bins between $t$ and $1.5\;t$ is plotted versus $t$.
It is seen that the slope for $t\lsim 10$ is not yet constant and therefore the
region $t<10$ should be excluded. It is also seen that the fluctuations within
the higher bins increase although much more points enter.
 
\vspace*{0.3cm}

Till now a completely disordered initial state has been considered as 
starting point, i.e., a state of very high
temperature. The question arises how a
{\em completely ordered initial state} evolves, when heated up suddenly to the
critical temperature. In the scaling form (\ref{e2}) one can skip
besides $L$, also the argument
$m_0=1$:
\begin{equation}
M^{(k)}(t,\tau) = b^{-k\beta/\nu} \;
           M^{(k)}(\;t/b^z,\;b^{1/\nu}\tau)\;\qquad (m_0=1).
\label{e7} 
\end{equation}

The system is simulated numerically by starting with a completely ordered state,
whose evaluation is measured at or near the critical temperature. The quantities
measured are $M(t)$ and $M^{(2)}(t)$. With $b=t^{1/z}$ one avoids 
the main $t$-dependence on the right of eq.(\ref{e7}), and for $k=1$ one has
\begin{equation}
M(t,\tau) = t^{-\beta/\nu z} \;
           M(\;1,\;t^{1/\nu z}\tau)\;
	   =t^{-\beta/\nu z}\left(1+a\;t^{1/\nu z}\;\tau+O(\tau^2)\right).
\label{e8} 
\end{equation}

\begin{figure}[h]
\begin{tabular}{cc}
\hspace*{-1cm}\raisebox{.5cm}{\includegraphics[height=4.5cm]{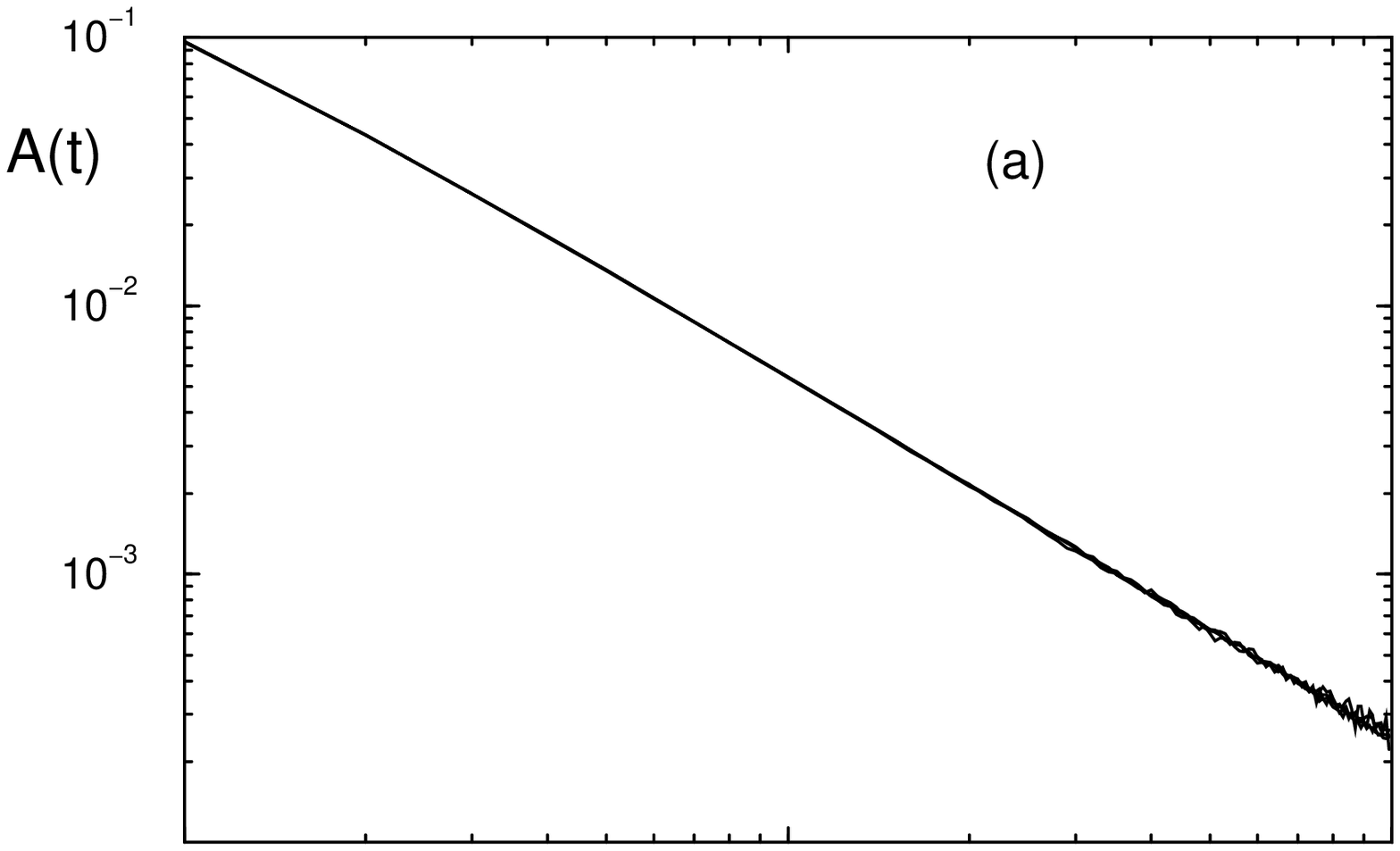}} 
&
\hspace*{-.6cm}\raisebox{.5cm}{\includegraphics[height=4.5cm]{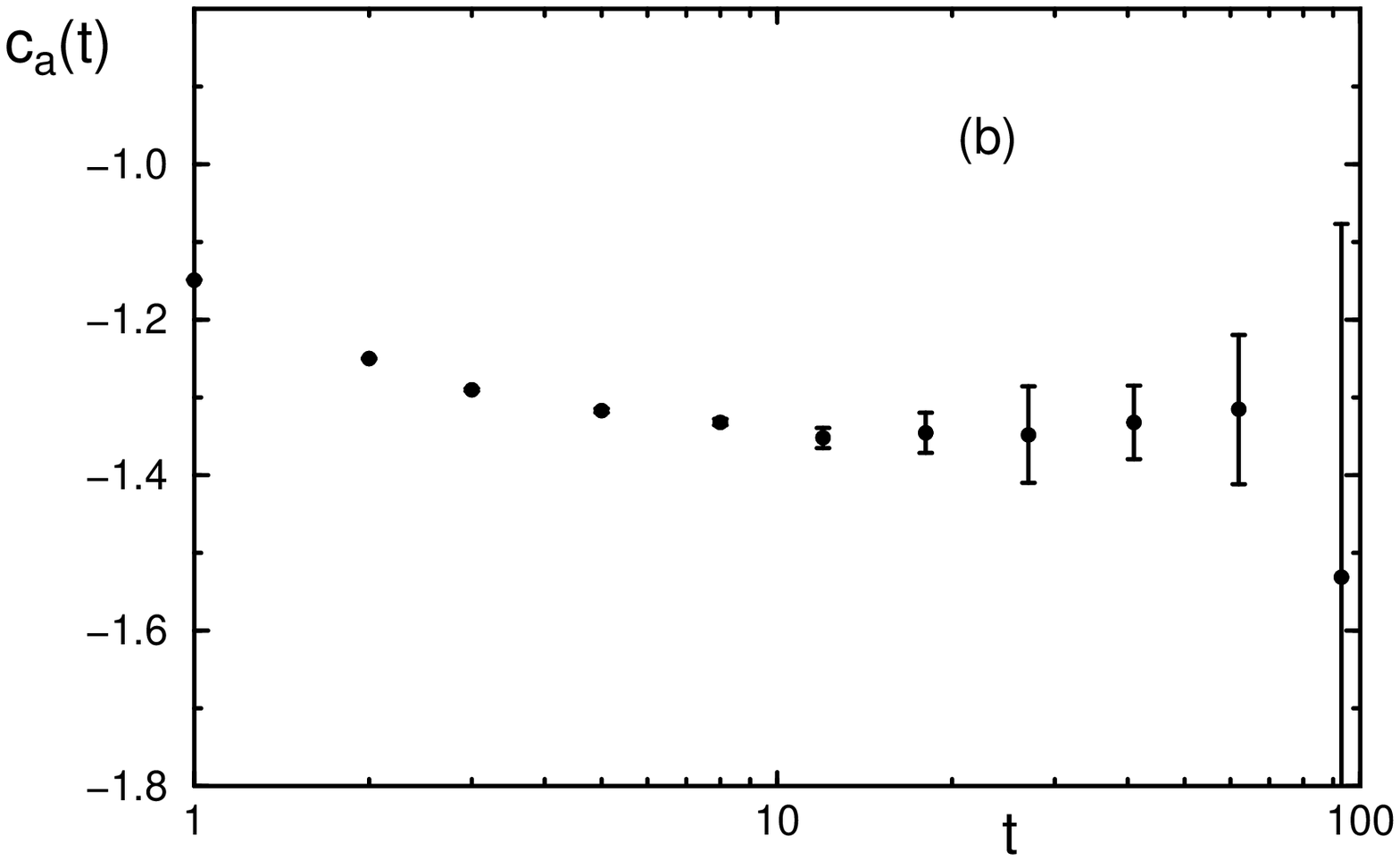}} \\ 
\end{tabular}
\end{figure}

\vspace*{-1.2cm}
\begin{tabular}{cc}
\hspace*{-1cm}\parbox[l]{5.7cm}{
Figure 5.
Three-dimensional Ising model:  Autocorrelation plotted vs. 
time on log-log scale. $L=128$. From the slope one gets $c_a=1.36(1)$}
&
\hspace*{0.5cm}{\parbox[r]{5.7cm}{Figure 6.
Three-dimensional Ising model: {slope $c_a$ } 
measured between 
${t}$ and ${1.5 t}$ plotted vs. 
time on log-log scale.}}
\\
\end{tabular}

\vspace*{.3cm}
Similarly as in case of the a completely disordered start, see eq.(\ref{e3}), 
one can choose couplings in the neighbourhood of the critical point and fix
from these measurements the critical point by looking for the best power law
behaviour. We would like to mention that measurements starting from $m_0=1$
are much less affected by fluctuations, because the quantities measured are
rather big in contrast to those from a random start. Figure~7 is an
example of the 3-dimensional Ising model\cite{jas99}. Both the critical point
as well as the slope at that point can be determined from the measurement.
Figure~8 shows the decay of the magnetisation at the critical point for
different lattice sizes from $L=32$ to $256$. The data points for all lattice
sizes agree completely up to $t=1000$, only the data for $L=32$ show some 
deviation for $t\gsim300$. This justifies our statement that finite size
effects are unimportant in short-time critical dynamics.

Also for the cold start one derives from (\ref{e7}) 
\begin{equation}
\partial_\tau\ln M(t,\tau)|_{\tau=0}\; \sim \; t^{{\;1}/{\nu z}}
\label{e9}
\end{equation}
which allows to measure the ratio ${1}/{\nu z}$. With the
magnetisation and its second moment  the time dependent Binder cumulant
\begin{equation}
U(t)=\frac{M^{(2)}}{(M)^2}-1\quad\sim\quad t^{d/z}
\label{e10}
\end{equation}
is defined. From its slope one can directly measure the dynamic exponent $z$.
For the last two quantities we give examples in Figs.~9 and 10,
using the 3-dimensional Ising model.

\begin{figure}[h]
\begin{tabular}{cc}
\hspace*{-1cm}\includegraphics[height=4.5cm]{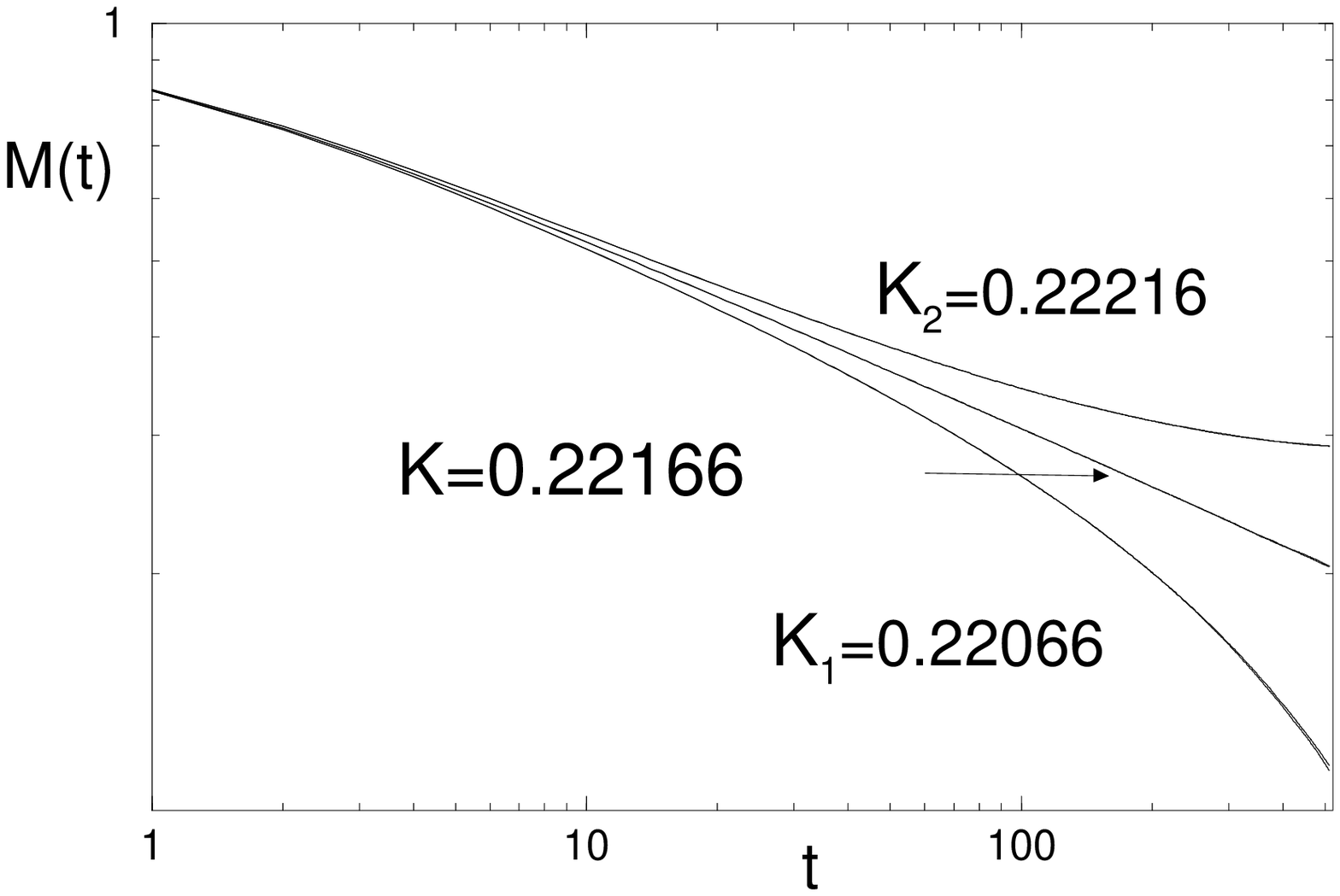} 
&
\hspace*{-.7cm}{\includegraphics[height=4.5cm]{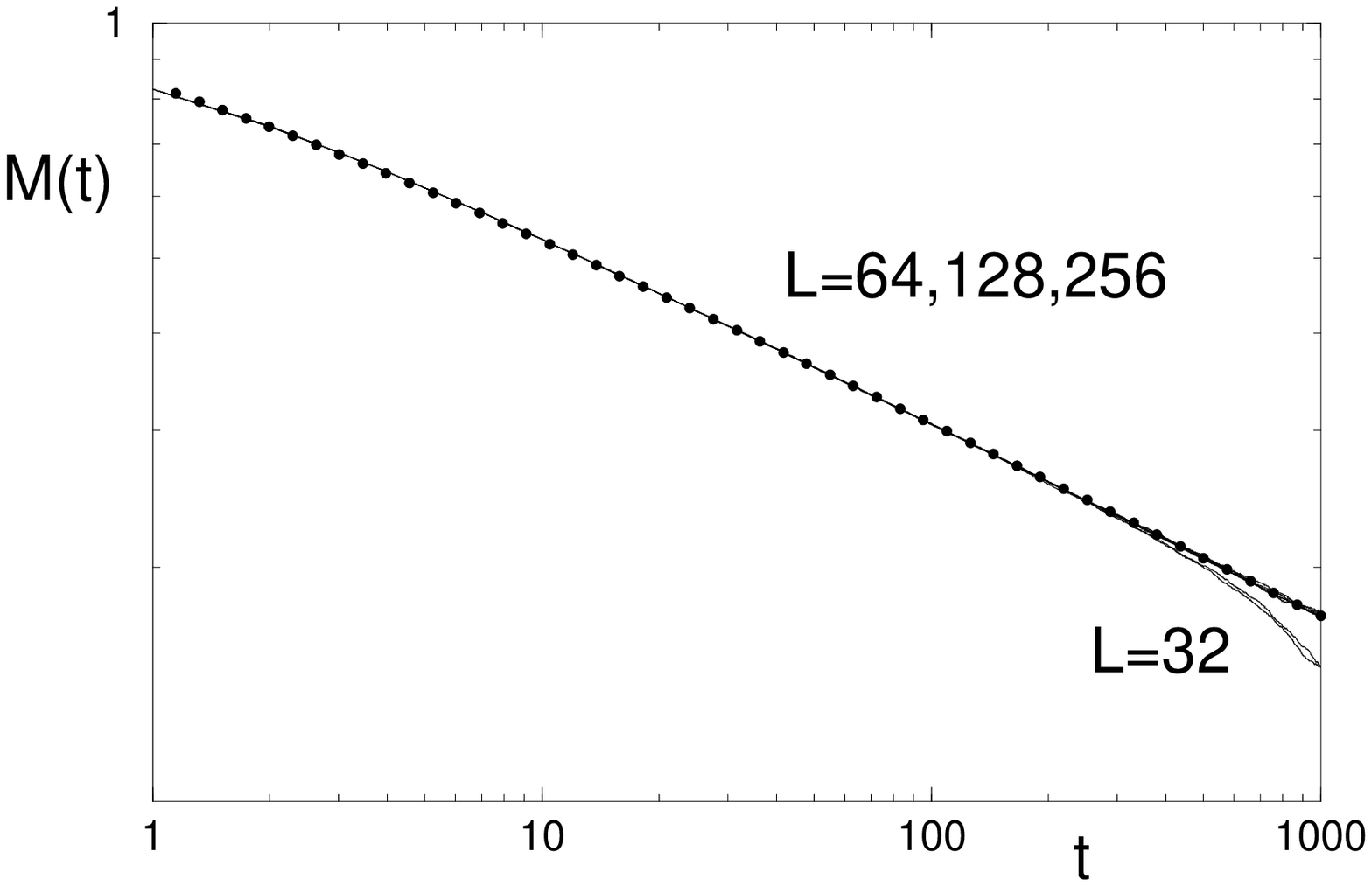}} \\ 
\end{tabular}
\end{figure}
\vspace{-.5cm}
\begin{tabular}{cc}
\hspace*{-0.9cm}\parbox[l]{5.7cm}{
Figure 7.
Three-dimensional Ising model: Decay of the 
magnetisation for initial magnetisations $m_0=1$, 
for three different values of the coupling near the critical point. }
&
\hspace*{0.2cm}\raisebox{.2cm}{\parbox[r]{5.7cm}{Figure 8.
Three-dimensional Ising model:  Decay of the magnetisation 
for different lattice sizes at the critical point.
}}
\\
\end{tabular}

\vspace*{0.4cm}
In summarising the topics discussed up to now, we would like to 
make the following remarks:
\begin{itemize}
\vspace*{-0.2cm}\item From an investigation of the system from a high-temperature initial 
state,
a new independent critical exponent $\theta$ can be determined.
\vspace*{-0.2cm}\item A determination of the critical point and of all the critical exponents
is possible starting from a high-temperature or from a zero-temperature
initial state. The results are the same for second-order transitions.
\vspace*{-0.2cm}\item In contrast to investigations in equilibrium, the correlation length is
still finite in the short-time regime, therefore finite size effects are strongly
reduced.
\vspace*{-0.2cm}\item Also the correlation time is still finite, therefore critical slowing
down is absent.
\vspace*{-0.2cm}\item We work directly with sample averages, not with time averages.
\end{itemize}

As is well known, critical slowing down can be overcome in some cases by the
successful non-local cluster algorithm\cite{swe87}. However, the cluster 
algorithm is not applicable, e.g., for systems with randomness or frustration. 
With short-time dynamic simulations we have, e.g., 
successfully investigated the chiral degree of freedom in the
2-dimensional fully frustrated XY model, where critical slowing  down makes
measurements in equilibrium extremely difficult\cite{luo98,luo98a}. 

\vspace*{0cm}
\begin{figure}[h]
\begin{tabular}{cc}
\hspace*{-1.2cm}\includegraphics[height=4.5cm]{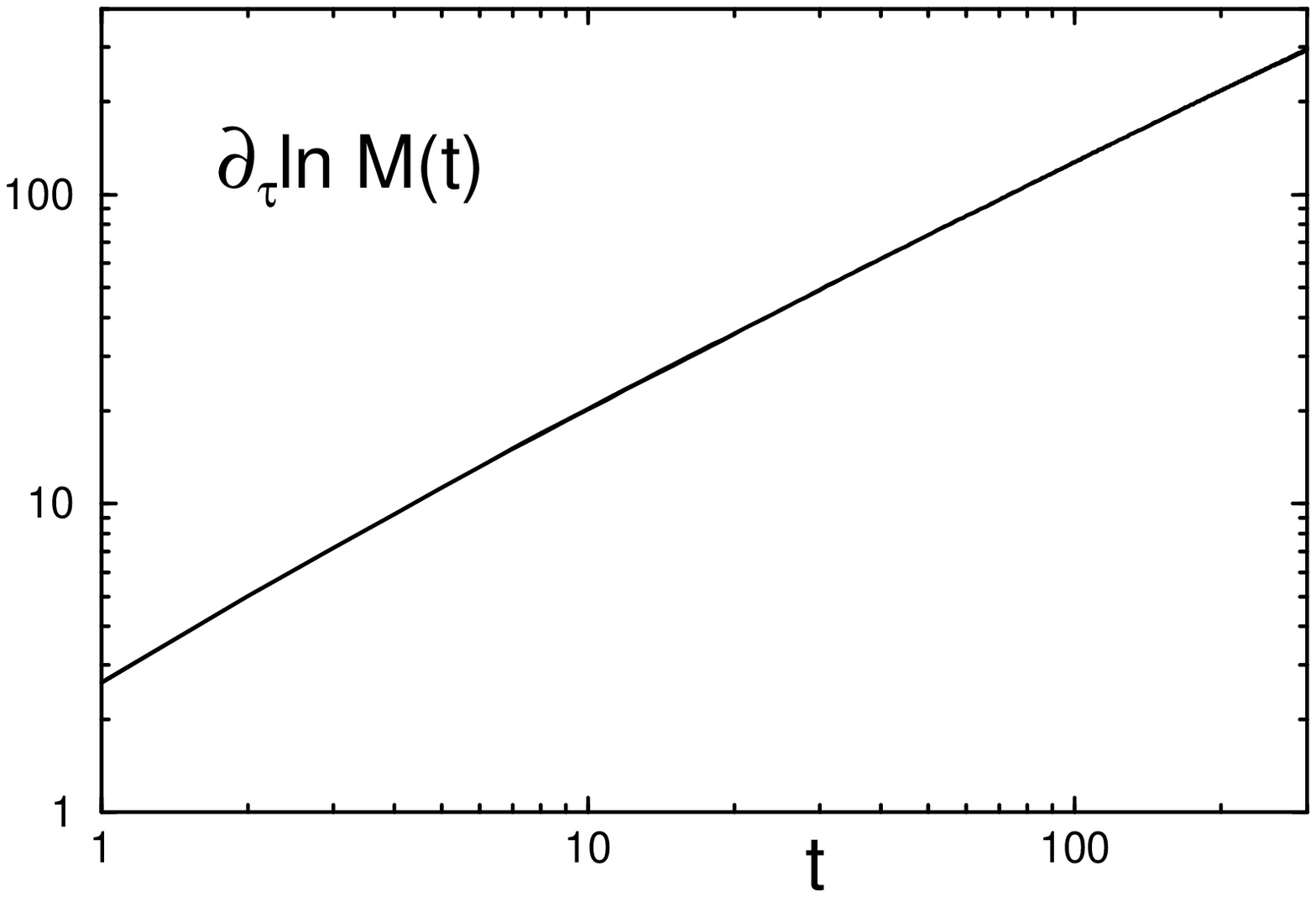} 
&
\hspace*{-.7cm}{\includegraphics[height=4.5cm]{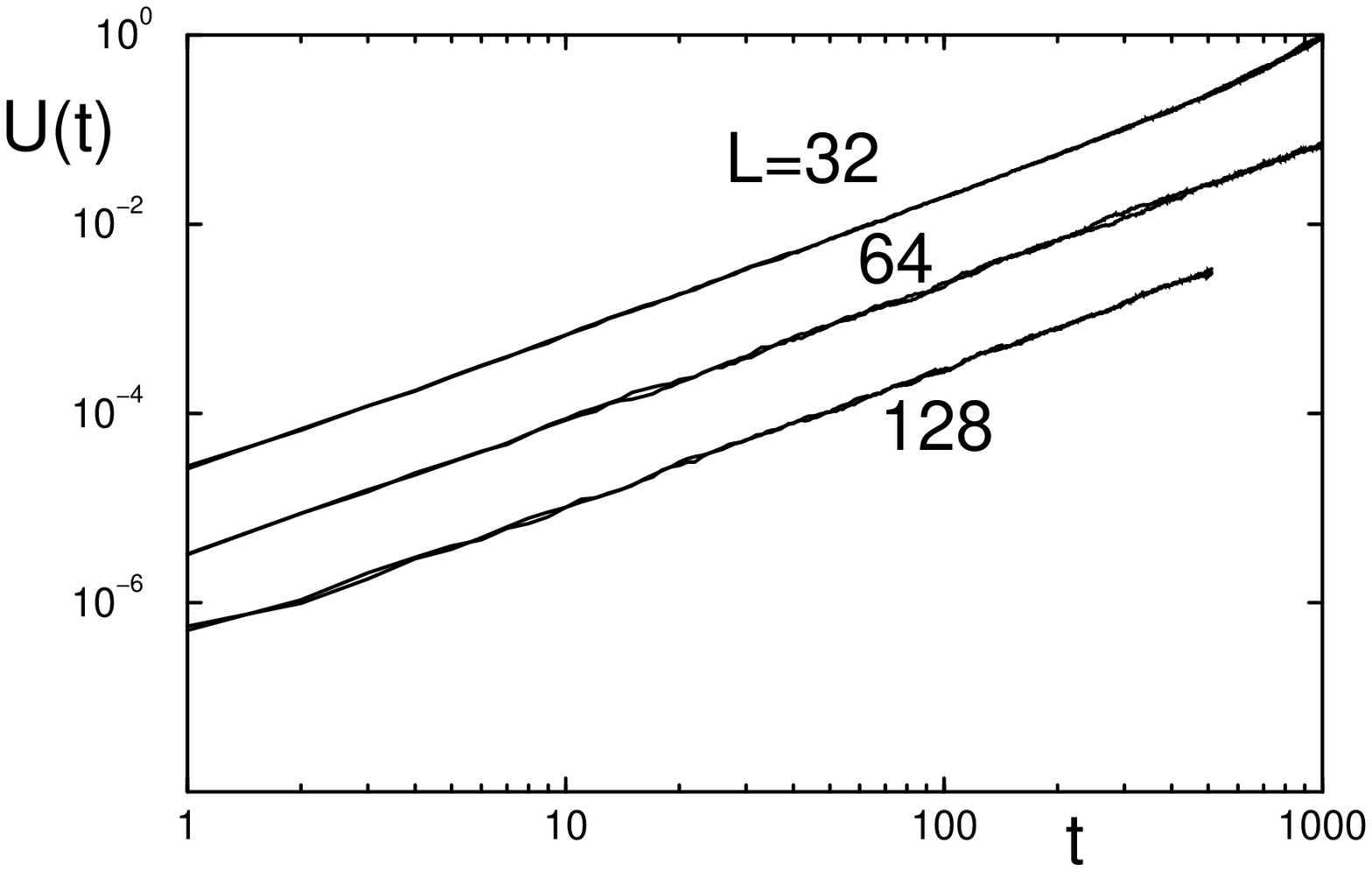}} \\ 
\end{tabular}
\end{figure}

\vspace*{-.9cm}
\begin{tabular}{cc}
\hspace*{-1cm}\parbox[l]{5.7cm}{
Figure 9. Three-dimensional Ising model: 
Logarithmic derivative of the magnetisation with respect to $\tau$,
obtained from an ordered start.}
&
\hspace*{0.2cm}\parbox[r]{5.7cm}{Figure 10. Three-dimensional Ising model: 
Time evolution of the Binder cumulant $U(t)$ for different lattice
sizes.}
\\
\end{tabular}

\vspace*{.6cm}

Till now we have investigated critical systems which undergo a second order
phase transition. The relaxation of initial states with very high temperature
suddenly quenched to the critical temperature, or of initial states with zero
temperature, suddenly heated up to the critical temperature, is measured in
the short-time regime. Various quantities exhibit a power law behaviour. From
measurements with couplings in the neighbourhood of the critical point we can
find by interpolation the optimal power law behaviour and determine 
in this way the critical
point. The result is the same for both initial conditions within the statistical
error.

If the phase transition is first order, at least if it is weak, the behaviour
should be similar to second order. By cooling down a very hot initial
state, one should find a {\em pseudo critical point $K^*$}, and by heating up a
cold initial state {\em another pseudo critical point $K^{**}$} is found. We
expect 
$$
K^{**} < K_c < K^*.
$$
A difference between $K^{**}$ and $K^*$ is a clear signal for a first-order
transition.

The p-state Potts model undergoes a second order phase transition for $p\leq4$
at the critical point $K_c=\ln(1+\sqrt p)$. For $p\geq5$ the transition is of
first order, but still weak for small $p$. We have investigated this
model\cite{sch00} for
$p=5$ and $p=7$. For the initial condition $m_0=0$ the second moment
$M^{(2)}(t)$ of the magnetisation has been measured, and for initial condition
$m_0=1$ the quantities $M(t)$ and $M^{(2)}(t)$. In case of $p=7$ the time
interval extended up to $t=6\;000$ for $L=280$ and $560$, while in case of 
$p=5$ a more careful analysis is necessary. Here we have chosen time intervals
up to $40\;000$ for $L=560$. In both cases we find a clear difference between
$K^*$ and $K^{**}$. In case of $K^{**}$, both the magnetisation and the Binder
cumulant (\ref{e10}) have been used. Figure~11 shows the result for $p=7$. The
results for $p=5$ are similar. This example shows that short-time critical
dynamics can be successfully used as a tool to distinguish between first- and second
order phase transitions. 

\vspace*{.3cm}\hspace*{1cm}\includegraphics*[height=7cm]{Ks_ss_7s.eps} 

\vspace*{0cm}
\hspace*{-0cm}\parbox[l]{12cm}{
Figure 11. 7-state Potts model: Determination of the pseudo-critical point
$K^{*}$ determined from $M^{(2)}(t)$ and of $K^{**}$ 
determined from $M(t)$ and $U(t)$. For the time interval $[t\cdots t_{max}]$
used for the fit various starting points between $t=100$ and $800$ and the 
maximum time have been used.
}


\end{document}